\documentclass[12pt]{article}
\usepackage{latexsym}
\usepackage{epsfig}
\newlength{\singlespace}
\newlength{\doublespace}
\newcommand{\etal}{{\em et al. }}
\newcommand{\beq}{\begin{equation}}
\newcommand{\eeq}{\end{equation}}

\title{Evidence for an extended scattered disk.}
\author{
 \\
\\
B.~Gladman$^1$, M.~Holman$^2$, T.~Grav$^3$,  \\
J.~Kavelaars$^4$, P.~Nicholson$^5$,
K.~Aksnes$^3$, J-M.~Petit$^1$
\\[3mm]
\\
$^1$Observatoire de la C\^ote d'Azur, France\\
$^2$Harvard-Smithsonian Center for Astrophysics, USA \\
$^3$University of Oslo, Norway \\
$^4$McMaster University, Canada \\
$^5$Cornell University, USA \\
\\
\\
\\
Proposed running head: Evidence for an extended  scattered disk.  \\
\\
\\
Submitted to Icarus, Mar 26/2001 
\\
\\
\\
22 text pages + 4 figures + 2 tables = 28 pages in initial submission
\\
}
\date{\rule{0mm}{0mm}}

\begin{document}
\maketitle

\setlength{\baselineskip}{\doublespace}

\newpage

\vspace*{2.5cm}

\begin{center}
{\bf Abstract}
\end{center}

\vspace*{2.5cm}


By telescopic tracking, we have established that the orbit of the 
trans-neptunian object (2000~CR$_{105}$) has a perihelion of $\simeq$44~AU,
and is thus outside the domain controlled by strong gravitational close 
encounters with Neptune.
Because this object is on a very large, eccentric orbit (with 
semimajor axis $a\simeq$216~AU and eccentricity $e\simeq$0.8)
this object must have been placed on this orbit by 
a gravitational perturbation which is {\it not} direct gravitational 
scattering off of any of the giant planets (on their current orbits).
The existence of this object may thus have profound cosmogonic implications
for our understanding of the formation of the outer Solar System. 
We discuss some viable scenarios which could have produced it, including
long-term diffusive chaos and scattering off of other massive bodies
in the outer Solar System.
This discovery implies that there must be a large population of 
trans-neptunian objects in an `extended scattered disk' with perihelia
above the previously-discussed 38~AU boundary.
 
\vspace{0.5cm}


\newpage
\section{Introduction}

Current nomenclature commonly divides the trans-neptunian region of the solar
system into (1) the `Kuiper Belt' (consisting of so-called `classical belt'
objects and `resonant objects' in various mean-motion resonances with
Neptune), and (2) the `scattered disk' (Jewitt \etal 1998).
Finer distinctions and sub-populations are possible (see 
Gladman \etal 2001).
The `scattered disk' is a structure that was observed to form naturally in 
simulations of (1) orbital perturbation of comets exterior to Neptune
(Torbett and Smoluchowski 1990) and (2) of the delivery of Jupiter family 
comets from Kuiper Belt (Duncan and Levison, 1997).  This latter work found
that as trans-neptunian objects (TNOs) leave the Kuiper Belt after
encountering Neptune, some are scattered outward to large, long-lived
external orbits rather than being passed inward to the other giant planets.
The first recognized member of this population of scattered disk objects 
(SDOs) was 1996~TL$_{66}$ (Luu \etal 1997); since that time of order
30 have been identified ({\it cf.}, Trujillo \etal 2000), with orbits
of varying quality. 
The semimajor axis, $a$, distribution of these
objects has no formal upper limit in the simulations of Duncan and
Levison (1997), although a distinction from the inner Oort cloud
becomes problematic at $a>1000$~AU (Duncan \etal 1987). 
The currently-known SDOs (Fig.~\ref{fig:alltnos}) must be concentrated towards
lower $a$ due to selection biases; larger-$a$ SDOs spend smaller
fractions of their time near perihelion where they appear brightest and
are more easily detected; a bias towards small perihelion distance $q$
will also exist.

Objects scattered to large orbits by Neptune will return
to near the planet at subsequent perihelion passages.  For a scattered object
the gravitational perturbation from Neptune might be considered as an
impulse as the object passes its perihelion; this alters the velocity
of the object but not its position.  
Thus, the $q$'s of scattered objects generally remain small, maintaining the
possibility of encounters with the planet.  Levison and Duncan
(1997, Fig.~6) show that except for some cases in the 2:1 resonance
with $a\simeq49~AU$, objects with $q>$40~AU are  entirely absent, or
at least are of extremely low probability.
Some objects have their perihelia raised from
35 to $\sim$39~AU due to a phenomena of `resonance sticking', to which 
we will return below.  
There is as yet no firm definition for the boundary between the SDOs
and Centaurs, nor even between the SDOs and the rest of the Kuiper
Belt (Gladman 2001).
Trujillo \etal (2000) seem to {\it define} the SDO population as that 
with $q$=34--36~AU.
We note in Figure 1 growing evidence for a $q\simeq$35--37~AU `gap' in
the SDO perihelion distribution, although statistics are still small
and assumptions in the orbits may still be important because many of the
SDO orbits are still not well sampled it time.

Once on scattered orbits, SDOs are subject to dynamical erosion as they 
are eventually perturbed by Neptune back onto orbits geometrically crossing 
that planet's orbit.  Then they are either (rarely) ejected by Neptune,
or have their perihelia pushed down to Uranus or below at which
point their dynamical lifetimes become $\sim$10~Myr.  At that point they are
usually rapidly removed from the solar system (Dones \etal 1996, 1997, Levison
and Duncan 1997). 
Nevertheless, the state with $q>30$~AU but still near Neptune can be very
long-lived due to the long orbital periods and the low probability of
Neptune being close when the object's rapid perihelion passage occurs.
Torbett (1989) and Torbett and Smoluchowski 
(1990) explored the long-term gravitational stability of such
large-$a$ orbits with $q$ near Neptune. They showed that such orbits
could be dynamically chaotic up to $q\simeq$45~AU, for large semimajor
axes. 
Although this chaotic evolution is not a sufficient condition for  
orbital instability ({\it cf.} Gladman 1993), it is in good agreement with
later long-term numerical integrations in the regime of common
exploration (out to about $a$=50~AU) (Duncan \etal 1995). 
However, the time scale for orbital instability is not established;
a chaotic orbit at high $a$ may require longer than 5~Gyr to reach
a state where it begins to interact strongly with Neptune.
Thus, large-$a$ orbits with $q>$38~AU are only weakly unstable to
the gravitational perturbations of the current giant planets over the 
lifetime of the solar system.
Later, we shall expand on some details of the above broad discussion in the
context of our observational result. 

\section{Observations}

The object 2000~CR$_{105}$ was discovered on 6 February 2000 in an
on-going survey by Millis \etal (2000), and, based on the observed 3-week arc
from a second night's observations on 27 February, the Minor Planet
Center (MPC) placed it on a provisional `scattered orbit' with 
$a$=82~AU, $e$=0.59, and $i$=31$^\circ$, implying $q$=33.8~AU.
The semimajor axis was chosen to be similar to that of 1996~TL$_{66}$
(B. Marsden, 2000, private communication).
Since the estimated heliocentric distance at the time of discovery was
$\simeq$55~AU, we realized this object was potentially of exceptional interest;
only the much fainter 1999 DG$_8$ (Gladman \etal 2001), 
at $62$~AU, had ever been discovered at such a large heliocentric distance.  
Thus, we re-observed 2000~CR$_{105}$ on 28 and 29 March 2000 at the
Canada-France Hawaii 3.5-m telescope;  given the short time interval
since the previous observation, we were stunned to find the object
already a dramatic 24 arcseconds off the ephemeris --- an enormous
positional error for a trans-neptunian object. This implied that the
object was moving eastward much more rapidly than 
indicated by the initial orbit and thus had a much larger semimajor axis.
Based on these observations, the orbit was revised by B.~Marsden to
$a$=675~AU, $e$=0.94, $i$=23$^\circ$, making it the largest scattered disk
orbit known.
Even with this preliminary orbit the perihelion ($q$=41~AU) had risen
out of the region believed to be strongly coupled to Neptune, but
given a two-month arc on an orbital period of greater than 10,000 yr,
this perihelion distance was still rather uncertain.  A further recovery
attempt by M.~Holman \etal in May 2000 at Kitt Peak failed in bad
weather, after which 2000~CR$_{105}$ disappered behind the Sun until
November 2000.

At this point there was a broad range of possible orbits for 2000~CR$_{105}$.
Its astrometry was still formally consistent, within observational
errors, with a parabolic orbit corresponding to a returning Oort cloud
comet (albeit with the most distant perihelion ever observed).
At the extreme end there was even the possibility that the orbit was
hyperbolic (corresponding to the first observed interstellar comet),
although this was less likely.  It was still possible that CR$_{105}$
would turn out to have $q<$39~AU and a  relatively `typical' scattered
disk orbit, but with a large semimajor axis.  Lastly, and most
interestingly, if a $q>$40~AU perihelion could be confirmed then, 
we believed at the time, 
this object would turn out to be the first SDO beyond the dynamical
influence of the giant planet system.  

We thus decided to allocate considerable telescope time to the recovery
and orbit determination of this object, beginning in the dark run
of November 2000 and continuing every dark run until February 2001, 
to provide a high-quality data set on which to base orbit calculations.
The first recovery was obtained with the European Southern Observatory's
2.2m telecope on 24 and 25~November 2000, followed a few days later by
confirming observations on the Hale 5~m telescope at Palomar.
During the following dark run at the Kitt Peak 4-m telescope, astrometry
was obtained on 17 and 18~December.
Astrometry on a single night was obtained on 20~January 2001 at the
Palomar 5-m, as well as on 16~February on the 2.5-m Nordic
Optical Telescope.  Lastly, the object was imaged on 23 and
24~February using the ESO VLT UT-1 telescope.
This brought the total observed arc to slightly more than one year, with 
extremely good time sampling in the recovery opposition.
It is worth noting that without our March 2000 observations this object
would have been 19 arcminutes away from its original ephemeris one year
after discovery; it likely would not have been recovered without considerable
effort, even with the large fields of view of mosaic cameras.  Based on this experience, it seems plausible that some of  
the TNOs that are observed for only short arcs in their discovery
opposition and then not found at their second opposition may very well
have orbits similar to that of 2000~CR$_{105}$.  We thus take a
fraction of 1 in $\sim$400 known TNO orbits as a lower limit to the
number of 2000~CR$_{105}$-like objects that have been detected in the
flux-limited TNO/SDO database. 

Based on it's apparent $R$-band magnitude of $m_R=23.3\pm0.5$, 
2000~CR$_{105}$'s diameter is roughly 400~km, based on an
assumed 4\% albedo.  We do not consider our photometric data reliable
enough, nor the assumed albedo accurate enough, to believe this to be
accurate to more than a factor of two.  This size places 2000~CR$_{105}$ at
the high end of the known range of trans-neptunian 
objects, about a factor of two below the largest known objects.

\section{Orbit modelling}

Using the available astrometric data, we have computed an osculating
orbit solution taking into account the perturbations of the 4 giant
planets (Table 1).
We used the orbit determination software developed by Berstein and 
Khushalani (2000), optimized specifically for outer Solar System 
objects.
This algorithm provides error estimates in the fitted osculating orbital
elements (Table 2).

Based on the available data, 2000~CR$_{105}$'s orbit is large, highly
elliptical, and moderately inclined (Table 2).  
Its semimajor axis exceeds that of any other currently known
multi-opposition TNO.  2000~CR$_{105}$ is currently 53~AU from the Sun
and moving outwards, having passed perihelion in mid-1965.  
At pericenter the object would have been about 0.8 magnitudes brighter.
The mean anomaly, argument of perihelion, and longtiude of node are all
well-determined.  
Overall this TNO might look like an outlier in the scattered disk 
distribution except for its very high $q=$44~AU perihelion (see Fig.~1). 
The perihelion distance has a fractional uncertainty much smaller than 
$a$ because the $a$/$e$ uncertainties are strongly correlated; 
the uncertainty in perihelion passage's distance is $<$1\%.

It is 2000~CR$_{105}$'s exceptionally large perihelion distance which 
merits special attention.
The only other SDO with a perihelion above 40~AU is 1995 TL$_8$ (Fig.~1), an
object discovered by the Spacewatch program (Larsen \etal 2001).

\section{Cosmogonic implications}

In this section we present several possible scenarios for how 2000~CR$_{105}$ 
arrived on its current orbit.  
These include perihelion raising by diffusive chaos, as well as scattering 
due to
(1) now-absent primordial embryos which passed through the forming Kuiper Belt, 
(2) a young Neptune that was forming a `fossilized scattered disk', or
(3) an unknown resident planetary-scale object in the distant Kuiper Belt.

\subsection{Diffusive chaos}

We have conducted a variety of numerical experiments to investigate the
long-term dynamics of 2000~CR$_{105}$.  In the first of these we
numerically integrated the best-fit orbit of this object, along with
20 other sets of initial conditions that are consistent with the
observations at the $1-\sigma$ level.  2000~CR$_{105}$ and its ``clones''
were modelled as test particles moving in the gravitational field of the giant
planets.  The planetary positions and velocities came from
the JPL DE403 ephemeris (with the terrestrial masses added to the
Sun).  
The integration algorithm was the symplectic $n$-body map of 
Wisdom and Holman (1991), with a time step of 0.5~year.
The integrations were for 5~Gyr of simulated time.

In addition to following the object trajectories, the tangent
equations for each trajectory were also integrated (Holman and Murray
1996, Mikkola and Innanen 1999).   This allows us to reliably
estimate the rate at which nearby trajectories diverge from each
other.  Regular or quasi-periodic trajectories separate from each other
linearly  or at most polynomially with time.  Chaotic motion is characterized
by exponential divergence of neighboring trajectories; the time scale of this
divergence is called the Lyapunov time.   

All of the trajectories share a number of characteristics:
 (1) Each is chaotic with a Lyapunov time from $5\times10^4$ to
$10^5$~years (15 to 30 orbital periods), consistent with previous studies 
of objects in this $a,e$ regime (Torbett 1989, Torbett and 
Smoluchowski 1990).  
 (2) the semimajor axis, $a$, oscillates rapidly within a series of discrete ranges.
 (3) The eccentricity, $e$, also varies rapidly; however, this variation is correlated 
with $a$ in such a way that $q$ varies much more smoothly.  
This occurs because the object receives an impulsive kick
from Neptune as it passes perihelion; at each conjunction the
position or perihelion distance of the object is nearly unaltered
but the velocity, and thus $a$ and $e$, is changed.  

Fig.~\ref{fig:evol1} shows a typical example of
the evolution of $a$ and $q$ for one test particle.   
The rapid $\sim$50~Myr
oscillation of $q$ is caused by the ``Kozai effect'' of Neptune on the
test particle (Kozai 1962), but its amplitude is far too low 
(see Thomas and Morbidelli 1996) to bring $q$ down to small values.
The center of each of the $a$-ranges, around which rapid oscillations
occur, corresponds to a high-order mean-motion resonance with Neptune.  
This demonstrates the so-called phenomenon of ``resonance sticking''.  
The resonance sticking seen proves that 2000~CR$_{105}$ is in or near
a regime in which chaotic phenomena are operating and that this region
of phase space may be connected to regions of lower perihelion by a 
an extended chaotic zone; this suggests the possibility 2000~CR$_{105}$ was on
a more `typical' SDO orbit which then diffused via chaotic phenomena to its
current high perihelion state.

The long time-scale variations of perhelion distance are 
important for evaluating the plausibility of this hypothesis.
Of the 20 integrated particles, 2 diffused to a minimum $q$ 
of 39~AU, although most remained between 42 and 47~AU 
(Fig.~\ref{fig:evolallq}).
We extended the integration of the test particle corresponding to the
best-fit orbital elements to determine the long-term fate of this object; 
the ``random walk'' in $q$ continued, with a maximum observed perihelion
distance of $q=50$~AU.  
An orbit with $q\sim40$~AU was attained after 24~Gyr, at which
point the semimajor axis diffused to very large ($10^3$~AU) values. 
Shortly thereafter $q$ dropped even lower, at which point
the particle was ejected from the solar system by Neptune.  
The integrations demonstrate that particles in the estimated orbital
region of 2000~CR$_{105}$ can, over very long time scales, reach perihelion
distances at which strong scatterings due to Neptune occur.  
Of course, the opposite can occur because the equations of motion are
time-reversible.   
The fact that none of the clones reached $q$=35~AU
on 5~Gyr time scales indicates that the probability of the reversed
process of going to a state near that of 2000~CR$_{105}$ is low.
In particular, the probability of leaving the vast chaotic zone to enter
the slowly-diffusing regime is unconstrained; the absence of such trajectories
in the Levison and Duncan (1997) simulations implies it is low.
We briefly note that the TNO 1996 TL$_8$ is near a which might plausibly
be reached by chaotic diffusion, but with a characteristic Lyapunov time
longer than 20 orbital periods.

\begin{figure}
\begin{center}
\epsfig{file=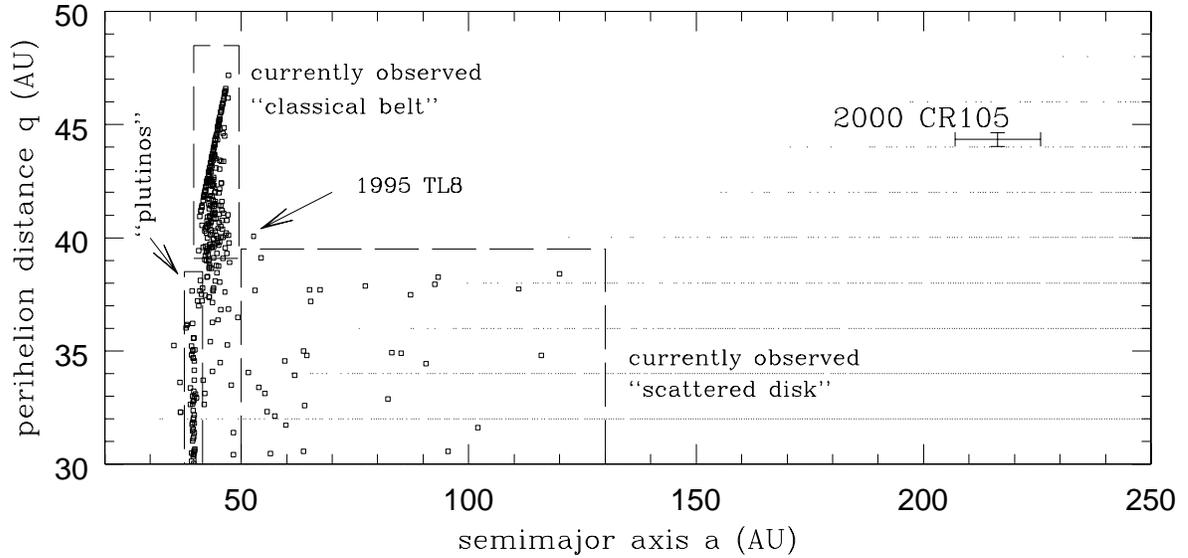,height=4.5in}
\end{center}
\caption{
Squares indicate estimated semimajor axes and perihelion distances for all 
trans-neptunian objects in the Minor Planet Center database as of 
February 2001.
Very approximate boundaries for membership in the classical Kuiper Belt, plutino
population, and scattered disk are indicated (only to guide the eye).
Small points indicate that a numerical integration (see text) showed a chaotic
orbital evolution for an orbit with those initial conditions.
Note that there are severe detection biases in this plot and it is not a
representative sampling of the trans-neptunian region.
\label{fig:alltnos}
}
\end{figure}

\begin{figure}
\begin{center}
\epsfig{file=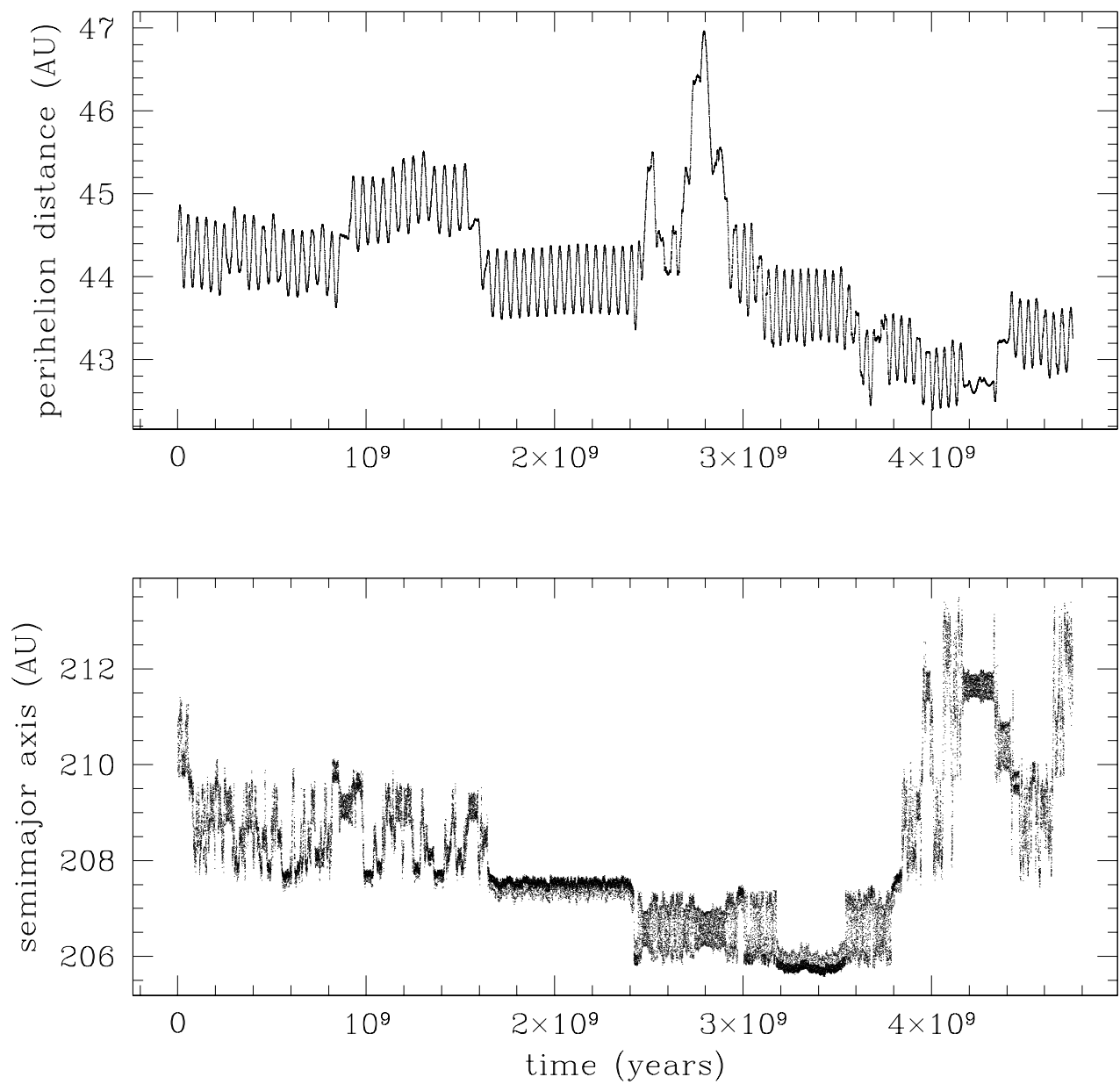,height=7.0in}
\end{center}
\caption{
The evolution for the semimajor axis $a$ and pericentric distance $q$
for a test particle integrated with an initial orbit consistent with that
of 2000~CR$_{105}$. See text for discussion.
\label{fig:evol1}
}
\end{figure}

\begin{figure}
\begin{center}
\epsfig{file=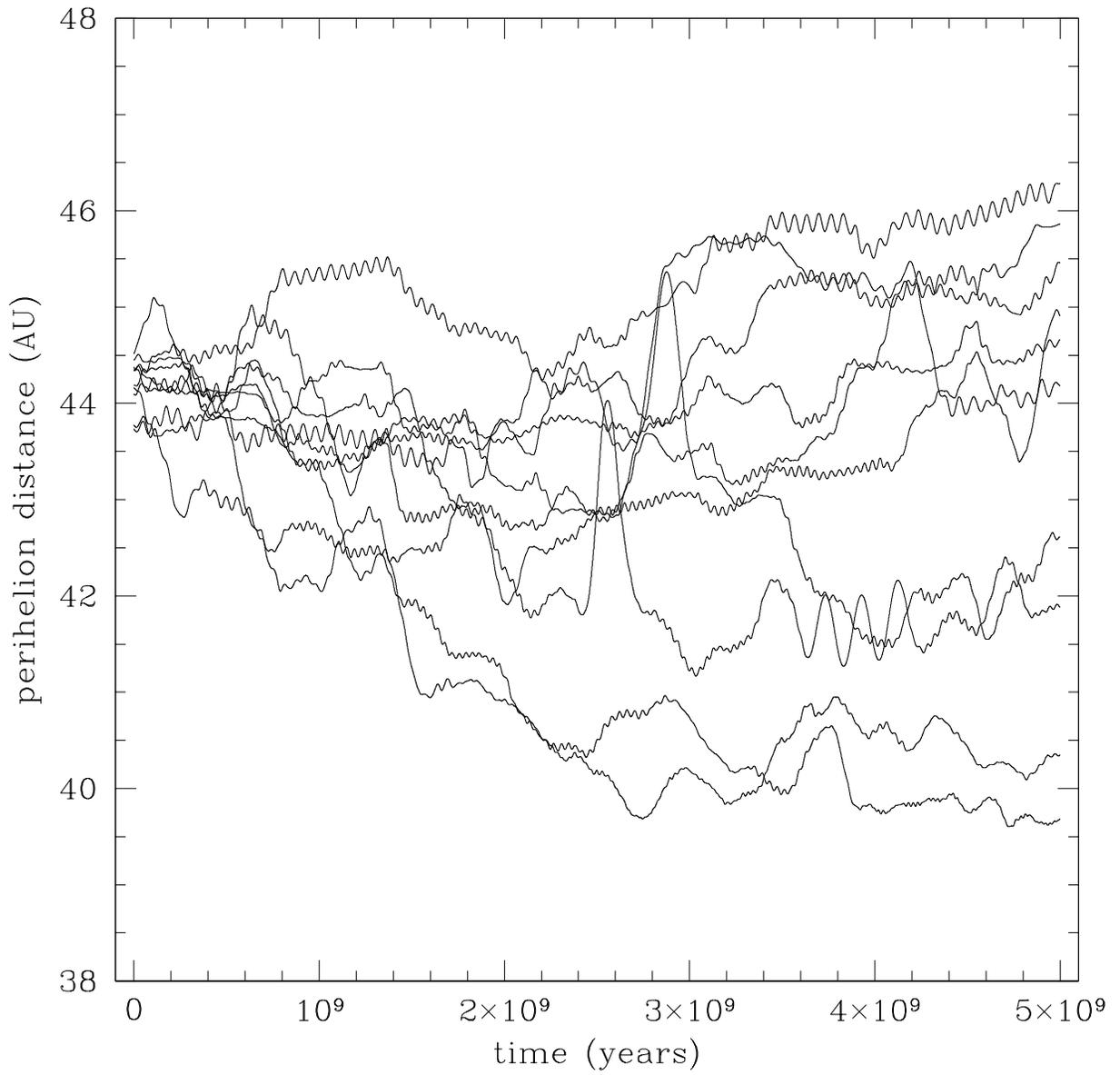,height=7.0in}
\end{center}
\caption{
The $q$-evolution for 10 of the 20 integrated `clones', showing
the range of variation for particles consistent with our best-fit orbit
for 2000~CR$_{105}$.
\label{fig:evolallq}
}
\end{figure}

\begin{figure}
\begin{center}
\epsfig{file=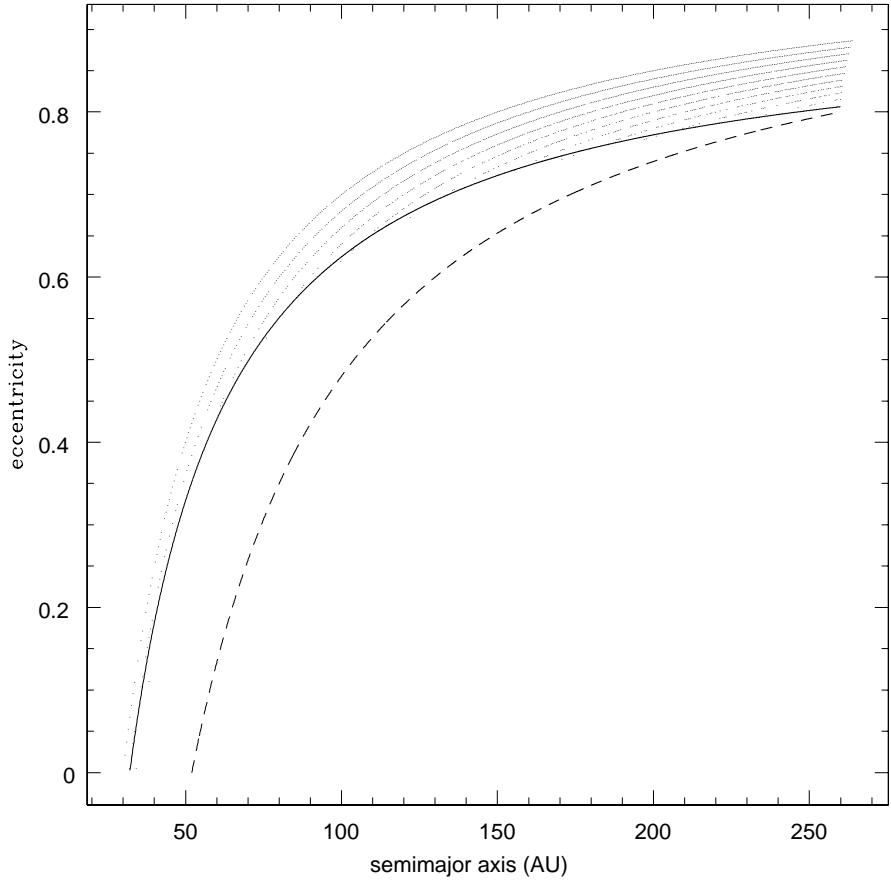,height=5.0in}
\end{center}
\caption{
An illustration of the chaotic structure of the region of large-$a$
solar system orbits.
Initial conditions were $a$=30--260~AU with $e$ selected to sample 
$q$=30--52~AU in increments of 2~AU.   
For the remaining elements we chose inclination $i=17^\circ$,
longitude of ascending node $\Omega=0^\circ$, 
argument of perihelion $\omega=0^\circ$, and mean anomaly $M=0^\circ$, 
with respect to the DE403 ecliptic and equinox. 
Each dot corresponds to an integrated trajectory (see text) with Lyapunov
time shorter than 20 of its orbital periods. 
The solid line denotes the relation 
$q=30 + 0.09(a-30)$ 
(determined empirically) which roughly bounds the envelope of
chaotic trajectories (also see Figure 1).  
Above this line, most of the trajectories exhibit strong, short time-scale 
chaos.  
The dashed line denotes the lower eccentricity boundary of the integrated
trajectories.  
Between the dashed line and the solid line few of the trajectories exhibited 
short time-scale chaos.  
The `fingers' of regular regions at large $a$ (absence of dots), reaching 
to higher values of $e$, correspond to the stable regions associated with 
individual mean-motion resonances.
\label{fig:ae}
}
\end{figure}

To accurately assess the probability of reaching the present orbit of
2000~CR$_{105}$ requires more extensive `forward' numerical integrations of
the formation of the scattered disk.
However, an associated issue is to determine the origin of the dynamical chaos
seen in the numerical integrations of 2000~CR$_{105}$ and its clones.  To do
so, we 
extended the work of Torbett (1989) and Torbett and Smoluchowski
(1990).   
We integrated 5400 test particles trajectories for $10^7$~years.   
We estimated the Lyapunov time of each trajectory, and based on their
histogram, we found that a value of 20~test-particle orbital periods 
separates those trajectories that are strongly chaotic from those that 
are not.  
As a test we checked that these high-$e$ test particles, when
integrated  without planetary perturbations, were not chaotic.
In Fig.~\ref{fig:alltnos} we plot a point at those values of $a$ and $q$
for which the corresponding trajectory was chaotic with Lyapunov time 
less than 20~periods.  
The envelope of these points is, surprisingly, nearly a straight, sloped
line.  Based on earlier descriptions of the boundaries of the scattered
disk chaotic zone, we expected to see chaos for trajectories with $q$ below 
a fixed value (a horizontal dividing line); our results show that SDOs
with large $a$ can have large $q$ and still exhibit chaos on
orbital-period time scales.  
In Fig.~\ref{fig:ae} we plot initial $a$ and $e$ of those trajectories 
with estimated Lyapunov times shorter than 20~orbital periods.  
The solid line corresponds to the envelope of chaotic
trajectories apparent on Fig.~\ref{fig:alltnos}.  
Few chaotic trajectories are found below the line.
The narrow `fingers' of non-chaotic trajectories that extend above the 
line correspond to the stable islands of high-order mean-motion 
resonances with Neptune.  
These resonances are narrow but not microscopically so;
at high $e$ resonance widths do not depend as strongly on the order
of the resonance as they do at low eccenticity.
An analytic estimate of the width
of the 6:1 mean motion resonance with Neptune, for example, yields
roughly 3~AU at Neptune-crossing eccentricity (Morbidelli \etal 1995).
On either side of the `fingers'
are chaotic regions resulting from the overlap of adjacent resonances.
A detailed resonance overlap calculation, extending the work of Wisdom (1980), 
can be completed at high eccentricity by employing the technique of 
Ferraz-Mello and Sato (1989).  

Malyshkin and Tremaine (1999) developed a 2-dimensional ``keplerian
map'', based on the planar restricted three body problem, to study the
long-term evolution of eccentric comet orbits perturbed by Neptune.
Although their mapping assumes a fixed perihelion distance and models the
entire gravitational interaction as an impulsive kick at perihelion
passage, their results capture many of the features seen in our direct
numerical integrations.  Two of their principle results are: (1) the
phase space is densely covered with chains of islands from mean-motion
resonances, and (2) the chaotic zone between these islands is
contiguous.  
That is, a trajectory can diffuse to arbitrarily large semimajor axis.  
Their first result we clearly see in our Fig.~\ref{fig:ae}.  
Their second result is apparent in the small perihelion values attained by
some of the 2000~CR$_{105}$ clones in our numerical integrations.

The details of our discussion can be refined once the semimajor axis of
2000~CR$_{105}$ is better determined, allowing us to determine exactly
where in phase space it resides; it could conceivably be in a dynamically more 
stable region although this seems unlikely.
Although we have given an extensive discussion of the diffusive chaos 
hypothesis (due to the fact that we can easily explore it numerically), the 
cosmogonic implication are even more dramatic if this hypothesis is 
either incorrect or untenable due to low probability. 
In such a case the existence of objects weakly coupled to the planetary system 
provides strong constraints on the formation of the outer Kuiper Belt.  
We now discuss three scenarios which could produce large numbers of such
weakly-coupled or decoupled TNOs.

\subsection{Primordial Embryos}

Given that 4 planets with masses greater than 10 Earth masses formed in the
outer Solar System, it is unlikely that no objects with martian-terrestrial
mass also formed in the region.
Morbidelli and Valsecchi (1997) and Petit \etal (2000) developed the idea that 
one or several of these objects (which they call `embryos') would have been
logically scattered outward by Neptune and spent some time as SDOs transiting the 
forming Kuiper Belt, thus causing the dynamical excitation and 
mass loss observed therein.
Close encounters with these passing embryos disturb the Kuiper belt out to the
aphelic distance of the embryos, which are often 50-100~AU; scattering events 
can thus produce TNOs with perihelia well past Neptune on high-$e$ orbits.
Once the embryos are eliminated by further gravitational interactions with
Neptune, as 90\% of SDOs are in 10 Myr (Duncan and Levison 1997), the
TNOs remaining are extremely long-lived.
In this scenario 2000~CR$_{105}$ may represent an object formed outstide of
50~AU which was scattered to a large-$e$ orbit due to encounters with one of 
these embryos.
Alternatively, 2000~CR$_{105}$ could have been formed much closer and {\it also}
became a SDO via scattering by Neptune and then an encounter with a 
distant embryo sufficed to raise $q$ to 44~AU.

\subsection{Fossilized scattered disk}

Thommes \etal (1999) propose that the `embryo' passing through the 
Kuiper Belt may have been Neptune itself, during a formation process
in which it transited the Kuiper belt on an orbit either more eccentric or 
with larger $a$ before reaching its present nearly circular orbit 
at 30~AU.
Being much more massive than the embryos discussed above, Neptune
would be
able to produce extensive dynamical `damage' in a shorter time. 
With its current $a$ and a  modest $e$ of $\sim$0.3, which Thommes \etal
damp via gravitational friction with a massive planetesimal disk in the
vicinity, Neptune's could encounter particles as far out as 44~AU.
After Neptune's aphelion evolves out of the 40~AU region (presumably rapidly)
the TNOs with $q$ above this limit are `fossilized' on orbits which either
do not evolve over the lifetime of the solar system, or evolve only slowly
via the diffusive mechanisms discussed above.
2000~CR$_{105}$ could be an example of the latter case.
Because the furthest $q$ of these objects will be just outside the largest
$Q$ of Neptune, the fossilized scattered disk forms an `arc' in $e$/$a$
space (see figures in Thommes \etal) separated from a `cold disk'
(see Gladman 2001) by a gap in $e$.  

\subsection{Resident planet}

A last scenario is that 2000~CR$_{105}$ arrived in its current dynamical state
due to gravitational interaction with a planetary-sized body that is 
{\it still resident} in the Kuiper Belt.
This could come about in two ways.  
First, in the distant Kuiper belt, beyond the region sculpted by the whatever
processes disturbed the 30-50~AU region, a planetary-mass body (the
size of the moon to Mars for example), or several, may have formed 
{\it in situ} over the lifetime of the solar system and the perturbations
of this body have sculpted the outer Kuiper belt.
A Mars-sized body (with escape velocity of 5~km/s) at 100~AU where orbital
velocities are only $\sim3$~km/s could scatter a body like 2000~CR$_{105}$
to its present orbit.

More likely may be a scenario in which several lunar---martian mass bodies 
were in the scattered disk, traversing in the 50--200~AU region.
Many such bodies were likely formed interior to Neptune as the cores of
the giant planets were accreting, and some would have ended up as SDOs.
Since orbital velocities at those distances are comparable to the escape
speeds of these bodies, mutual encounters between the `embryos' could
place one or more of them on orbits entirely decoupled with
Neptune; the planetary embryos still coupled to Neptune would have been
rapidly destroyed, leaving the decoupled embryo(s) `lodged' in the distant
Kuiper belt.
The resident embryo would then proceed to excite the orbital distribution,
over the age of the Solar System, of the entire Kuiper Belt between its
perihelion and aphelion distance.
Similar ideas trace back to Fern\'andez (1980) and Ip (1989), who sought
to push short-period comets to Neptune-crossing orbits via decoupled
embryos before long-term gravitational erosion was characterized as a 
supply process (Levison and Duncan 1997); we view the resident embyro
scenario as potentialy providing the ability produce objects like 
2000~CR$_{105}$ and to remove many objects beyond the 2:1 resonance.
Unpublished simulations by Morbidelli (2000, private communication) and
Brunini (1999, private communication) show that this decoupling and 
`clearing out' process can happen naturally. 
This scenario could be responsible for the apparent lack of objects on 
nearly circular orbits outside the 2:1 resonance 
(see Jewitt \etal 1998, Allen \etal 2000, and Gladman \etal 2001 for
discussion) and the lack of detection of the so-called `Kuiper Wall'
(Trujillo 2000).

\section{Conclusion}

We estimate that the majority of the TNO surveys to date, searching within
$\sim 5^\circ$ of the ecliptic to limiting magnitude $m_R$=23--24, are
capable of detecting 2000~CR$_{105}$ for $<1$\% of its orbital period.  Given
this extreme detection bias against finding objects like 2000~CR$_{105}$,
there must be a large number of objects with perihelia higher than the
34--36~AU range for the scattered disk (used by Trujillo \etal 2000).

Thus, the `scattered disk' may be much more massive than that
component which is currently strongly coupled to Neptune, and likely
merges into an `extended scattered disk' where objects are only weakly
coupled to Neptune.
If the eccentricity and inclination distribution does not have the structure
of a fossilized disk (with a gap in eccentricity) then it will be difficult 
to say where the `extended scattered disk' ends!
We propose that for the moment this object should {\it not} be classified as 
a scattered disk object unless a definition can be arrived at
which would delineate SDOs from an object on a regular orbit (for example,
$a$=210 and e=0.7, which is neither chaotic nor coupled to Neptune, 
and yet clearly not a member of the cold disk.
These nomenclature problems are expanded upon in Gladman (2001).

The lesson provided by our tracking of this exceptional TNO is that 
follow up inside the first year (2--3 months after discovery) is critical
in order to detect these large orbits, for without such a recovery the
TNO will be very far from even a `normal' scattered orbit one year later, at
which point recovery becomes difficult.
It is difficult to estimate what fraction of previously-lost objects may
have been similar to 2000~CR$_{105}$;
of the objects {\it already detected in the flux-limited sample},
(roughly 30 SDOs and/or $\sim$400 TNOs designated), the existence of
2000~CR$_{105}$ likely implies a strong lower limit of 5\% of SDOs and
$\sim$0.2\% of TNOs to be on orbits like 2000~CR$_{105}$, given the fraction
of the TNOs that have had observations on only a single opposition.

We are not yet able to estimate the likelihood that 2000~CR$_{105}$ has 
diffused to a large $q$ orbit if it was scattered to $a\sim$200~AU by
Neptune.
The rarity of similar particles in the integrations of Levison and
Duncan (1996) may simply mean that this is rarer than $\ll$1 in $10^3$ of
the SDOs initially populating the scattered disk; the fraction of the
SDOs {\it remaining} in this state after 5~Gyr will likely be roughly
two orders of magnitude higher (Duncan and Levison 1996) and thus only
1 in $10^5$ initial SDOs may need to reach the high-$q$ state.
However, given the direct detection bias against, and the possible earlier loss
of other already-discovered objects that may have been on similar orbits,
the actual fraction of the visible trans-neptunian region that is also
on high-$q$ orbits could be as large as several percent.
In this case the diffusive hypothesis may become untenable and some of the
more cosmogonically dramatic scenarios may be necessary.
If a high-$a$ TNO totally decoupled from the planetary system can be 
identified then the reality of these dramatic scenarios can be constrained;
the fact that the first high-$q$ TNO is in the diffusive boundary regime would
then be understood to be due to detection bias which favors the latter's
discovery.
One cannot stress enough that continued tracking of {\it all} discovered 
TNOs, especially 2--3 months after discovery, is necessary in order 
to insure that additional interesting objects are not lost.

\section{Acknowledgements}

B.~Gladman, J.~Kavelaars, and M.~Holman were visiting astronomers at 
the Canada France Hawaii telescope, operated by the National Research 
Council of Canada, le Centre National de la Recherche Scientifique de 
France, and the University of Hawaii.
Data from VLT collected at the European Southern Observatory, Chile, 
proposals 66.C-0048A and 66.C-0029A.
Observations at the Palomar Observatory were made as part of a continuing 
collaborative agreement between the California Institue of Technology
and Cornell University. 

K.~Aksnes, T.~Grav, and M.~Holman were visiting astronomers at the Nordic
Optical Telescope.  The Nordic Optical Telescope is operated on the island of
La Palma jointly by Denmark, Finland, Iceland, Norway, and Sweden, in the
Spanish Observatorio del Roque de los Muchachos of the Instituto de
Astrofisica de Canarias.  Some of the data presented here have been taken
using ALFOSC, which is owned by the Instituto de Astrofisica de Andalucia
(IAA) and operated at the Nordic Optical Telescope under agreement between IAA
and the NBIfAFG of the Astronomical Observatory of Copenhagen. 

M.~Holman, B.~Gladman, and T.~Grav were visiting astronomers at the Kitt Peak
National Observatory, National Optical Astronomy Observatory, which is
operated by the Association of Universities for Research in Astronomy,
Inc. (AURA) under cooperative agreement with the National Science Foundation.

This work was supported in part by an ACI Jeune grant from the French
Ministry of Research, and NASA grants NAG5-10365 and NAG5-9678 to the
Smithsonian Astrophysical Observatory.

We thank Joe Burns, Brian Marsden, Alessandro Morbidelli, and Norm Murray 
for helpful input.

\newpage

\noindent
{\bf References}\\
\noindent
\setlength{\parindent}{-5pt}
\quad

Allen, L., Bernstein, G., and Malhotra, R. 2001.
The edge of the solar system.
{\it Ap. J. Lett.} {\bf 549}, 241--244.

Bernstein, G. and Khushalani, B.  2000. 
Orbit Fitting and Uncertainties for Kuiper Belt Objects.
{\it A. J.}{\bf 120}, 3323-3332.

Dones, L., H.~F. Levison, and M.~Duncan 1996. On the dynamical
lifetimes of planet-crossing objects. In {\it Completing the
Inventory of the Solar System} (T.~W. Rettig, J.~Hahn Eds.), Astronomical
Society of the Pacific Press, Vol.~107, pp.~233--244.

Dones, L., Gladman, B., Melosh, H. J., Tonks, W. B., Levison, Harold F.,
Duncan, M. 1997.
Dynamical Lifetimes and Final Fates of Small Bodies: Orbit Integrations vs
Opik Calculations.
{\t Icarus} {\bf 142}, 509-524.

Duncan, M., Quinn, T., Tremaine, S. 1987. 
The formation and extent of the solar system comet cloud.
{\it A. J.} {\bf 94}, 1330-1338. 

Duncan, M. J., Levison, H. F. (1997)
A scattered comet disk and the origin of Jupiter family comets.
{\it Science} {\bf 276}, 1670-1672.

Ferraz-Mello, S., Sato, M. 1989.
The very-high-eccentricity asymmetric expansion of the disturbing
function near resonances of any order. 
Astronomy and Astrophysics {\bf 225}, 541--547.

Gladman, B. 2001. Nomenclature in Kuiper Belt,
in {\it Proceedings of Joint Discussion 4 of the 2000 IAU General Assembly}. 
(A. Lemaitre Ed.) Kluwer, in press.

Gladman, B., Kavelaars, J., Petit, J-M., Morbidell, A., Holman, M., Loredo, T.
2001.
The Structure of the Kuiper Belt: Size Distribution and Radial Extent.
A.J., submitted.

Jewitt, D. G., Luu, J., and Chen, J. 1996. 
The Mauna Kea-Cerro-Tololo Kuiper Belt and Centaur Survey. 
{\it A. J.} {\bf 112}, 1225-1238.

Jewitt, D. G., Luu, J., and Trujillo, C. 1998.
Large Kuiper Belt objects: The Mauna Kea 8K CCD survey. 
{\it A. J.} {\bf 115}, 2125-2135.

Larsen, J. A., Gleason, A. E., Danzl, N. M., Descour, A. S., McMillan,
R. S.; Gehrels, T., Jedicke, R.,  Montani, J. L., Scotti, J. V.
2001. 
The Spacewatch Wide-Area Survey for Bright Centaurs and Trans-Neptunian Objects.
{\it Astron. J.} {\bf 121}, 562-579.

Levison, H. F., Duncan, M. J. 1997.
From the Kuiper Belt to Jupiter-Family Comets: The Spatial
Distribution of Ecliptic Comets. 
{\it Icarus}, {\bf 127}, 13-32.

Malyshkin, L., Tremaine, S. 1999.
The Keplerian Map for the Planar Restricted Three-Body Problem as a Model of Comet Evolution.
Icarus, {\bf 141}, 341--353.	

Mikkola, S., Innanen, K. 1999.
Symplectic Tangent map for Planetary Motions
{\it Celest. Mech. Dyn. Astron.} {\bf 74}, 59--67.

Millis, R. L., Buie, M. W., Wasserman, L. H., Elliot,
J. L., Kern, S. D., Wagner, R. M. 2000.
The Deep Ecliptic Survey.
{\it B.A.A.S.} {\bf vol}, ??

Morbidelli, A., Thomas, F., Moons, M. 1995.
The resonant structure of the Kuiper belt and the dynamics of the
first five trans-Neptunian objects. 
Icarus, {\bf 118}, 322--340.

Morbidelli, A. and Valsecchi, G. 1997, 
NOTE: Neptune Scattered Planetesimals Could Have Sculpted the Primordial Edgeworth-Kuiper Belt.
Icarus, {\bf 128}, 464.

Petit, J-M., Morbidelli, A. and Valsecchi, G. 1999.
Large Scattered Planetesimals and the Excitation of the Small Body Belts.
Icarus, {\bf 141}, 367.

Torbett, M. 1989. 
Chaotic motion in a comet disk beyond Neptune: 
   The delivery of short-period comets.
A.J. {\bf 98} 1477-1481.

Torbett, M. and Smoluchowski, R. 1990.
Chaotic motion in a primordial comet disk beyond Neptune and comet inlux
  to the solar system. 
Nature, {\bf 345}, 49--51.

Thomas, F., and A.~Morbidelli (1996) 
The Kozai resonance in the outer solar system and the dynamics of 
long--period comets
Celest.  Mech., {\bf 64}, 209--229.

Thommes, E., Duncan, M., Levison, H.F. 1999.
The formation of Uranus and Neptune in the Jupiter-Saturn region
of the Solar System.
Nature, {\bf 402}, 635--638.

Trujillo, C. 2000.
Simulations of the bias effects in Kuiper Belt surveys.
in {\it Minor Bodies in the Outer Solar System} (A.~Fitzsimmons,
D.~Jewitt, R.M.~West, Eds.) pp. 109--115.  Springer-Verlag, Berlin.

Trujillo, C., D. Jewitt, Luu, J. 2000.
Population of the Scattered Kuiper Belt
{\it Ap. J.} {\bf 102}, 529--533.

Wisdom, J. 1980. 
The resonance overlap criterion and the onset of stochastic behavior 
in the restricted three-body problem.
{\it Astron. J} {\bf 85}, 1122-1133.

Wisdom, J., Holman, M. 1991.
Symplectic maps for the n-body problem.
{\it A. J.} {\bf 102}, 1528--1538.

\setlength{\baselineskip}{\singlespace}



\newpage

\pagestyle{empty}


{\bf Table 1. Astrometric observations and calculated orbit for 2000 CR$_{105}$}

\vspace*{0.4cm}

\begin{small}

\begin{tabular}{lcccccl}
\hline
{Object} & {UT Date} &{$\alpha$(2000) }
&{$\delta$ (2000)} & {R} & {Obs.} &{Note}\\
& {yyyy~mm~dd.ddddd} &{hh~mm~ss.ss} & {~dd~mm~ss.s} & {mag} & {Code}  \\
\hline
K00CA5R & 2000~02~06.30637 &09~14~02.39 &+19~05~58.7 &22.5 R  & 695 &  1 \\
K00CA5R & 2000~02~06.43541 &09~14~01.90 &+19~06~01.4 &        & 695 &  1 \\
K00CA5R & 2000~02~27.12907 &09~12~44.37 &+19~13~04.6 &23.0 R  & 695 &  1 \\ 
K00CA5R & 2000~02~27.22612 &09~12~43.98 &+19~13~06.3 &        & 695 &  1 \\
K00CA5R & 2000~03~28.38346 &09~11~17.68 &+19~20~37.4 &        & 568 &  2 \\
K00CA5R & 2000~03~28.40927 &09~11~17.63 &+19~20~37.6 &        & 568 &  2 \\
K00CA5R & 2000~03~28.43164 &09~11~17.59 &+19~20~37.9 &        & 568 &  2 \\
K00CA5R & 2000~03~29.23055 &09~11~15.99 &+19~20~46.3 &        & 568 &  2 \\
K00CA5R & 2000~03~29.25196 &09~11~15.95 &+19~20~46.5 &23.1 R  & 568 &  2 \\
K00CA5R & 2000~03~29.27248 &09~11~15.91 &+19~20~46.7 &23.4 R  & 568 &  2 \\
K00CA5R & 2000~11~24.30080 &09~22~07.39 &+18~49~14.6 &        & 809 &  3 * \\
K00CA5R & 2000~11~25.30941 &09~22~06.97 &+18~49~25.1 &        & 809 &  3 * \\
K00CA5R & 2000~11~27.48283 &09~22~05.77 &+18~49~49.1 &23.5 R  & 675 &  4 \\
K00CA5R & 2000~11~27.54557 &09~22~05.73 &+18~49~49.1 &23.7 R  & 675 &  4 \\
K00CA5R & 2000~11~28.48968 &09~22~05.11 &+18~49~59.8 &24.1 R  & 675 &  4 \\
K00CA5R & 2000~11~28.52381 &09~22~05.08 &+18~50~00.3 &23.6 R  & 675 &  4 \\
K00CA5R & 2000~12~17.45692 &09~21~38.68 &+18~54~34.5 &23.0 R  & 695 &  5 \\
K00CA5R & 2000~12~17.49240 &09~21~38.62 &+18~54~34.9 &23.2 R  & 695 &  5 \\
K00CA5R & 2000~12~17.53331 &09~21~38.54 &+18~54~35.9 &23.4 R  & 695 &  5 \\
K00CA5R & 2000~12~18.43274 &09~21~36.63 &+18~54~51.1 &23.0 R  & 695 &  5 \\
K00CA5R & 2000~12~18.50613 &09~21~36.52 &+18~54~52.2 &22.8 R  & 695 &  5 * \\
K00CA5R & 2001~01~20.39079 &09~19~58.51 &+19~06~04.4 &23.1 R  & 675 &  6 \\
K00CA5R & 2001~01~20.39910 &09~19~58.46 &+19~06~04.6 &        & 675 &  6 \\
K00CA5R & 2001~01~20.46716 &09~19~58.24 &+19~06~06.2 &23.1 R  & 675 &  6 \\
K00CA5R & 2001~02~15.99209 &09~18~17.96 &+19~15~46.4 &23.7 R  & 950 &  7 \\
K00CA5R & 2001~02~16.03317 &09~18~17.80 &+19~15~47.2 &23.4 R  & 950 &  7 \\
K00CA5R & 2001~02~16.08746 &09~18~17.59 &+19~15~48.4 &23.5 R  & 950 &  7 \\
K00CA5R & 2001~02~23.15703 &09~17~51.48 &+19~18~11.2 &23.3 R  & 309 &  8 \\
K00CA5R & 2001~02~24.24770 &09~17~47.52 &+19~18~32.4 &        & 309 &  8 \\
\hline
\end{tabular}

\vspace*{4.0mm}

Notes: 1 -- Millis {\it et al.} KPNO 4-m and WIYN (MPEC 2000-F07) \\

\vspace*{-4.9mm}
2 -- Gladman, Kavelaars, Holman, Petit (CFHT-3.5m); 3 -- Gladman (ESO-2.2m); \\

\vspace*{-4.9mm}
4 -- Nicholson, Kavelaars (Palomar-5m); 5~--~Holman, Gladman, Grav (KPNO-4m);\\

\vspace*{-4.9mm}
6 -- Nicholson, Gladman (Palomar-5m); 7 -- Grav, Holman (NOT-2.5m);\\ 

\vspace*{-4.9mm}
8 -- Gladman (VLT UT1-8m)\\ 

\vspace*{-2.9mm}
Astrometric uncertainties~$\alpha \pm 0.03 s$ , $\delta \pm 0.4^{\prime\prime}$
except {\tt *} for which $\alpha \pm 0.07 s$, $\delta \pm 1^{\prime\prime}$. \\

\vspace*{-5.9mm}
Photometric uncertainties $\pm$~0.5 mag
\end{small} 

\vspace*{0.3cm}

\newpage

\begin{large}

\begin{center}

{\bf Table 2. Orbital elements for 2000 CR$_{105}$}

\vspace*{0.3cm}

\begin{tabular}{lcc} \hline
Orbital element (J2000)         & Value         & 1-sigma error \\ \hline
semimajor axis $a$        	& 216 AU	& 9 AU \\
eccentricity $e$	        & 0.795		& 0.010 \\ \hline
{\bf perihelion distance} $q$	& 44.2	  	  & 0.3 AU \\ \hline
inclination $i$		        & 22.759$^\circ$  & 0.002$^\circ$  \\
longitude of node $\Omega$      & 128.287$^\circ$ & 0.001$^\circ$  \\
argument of pericenter $\omega$ & 317.0$^\circ$   & 0.6$^\circ$  \\
Date of pericenter passage (JD) & 2438870         & 70 \\ 
mean anomaly $M$	        & 3.9$^\circ$     & 0.2$^\circ$ \\ \hline
\end{tabular}
\end{center}

\end{large}

\newpage

\end{document}